%% file: main.tex
\pdfoutput=1
\documentclass[11pt]{article}

\usepackage[final]{acl}
\usepackage{amsmath} 

\usepackage{graphicx}
\usepackage{multirow}

\usepackage{url}
\usepackage{booktabs}
\usepackage{amssymb}
\usepackage{bbding}
\usepackage{pifont}
\usepackage{wasysym}
\usepackage{times}
\usepackage{latexsym}
\usepackage[ruled,vlined,linesnumbered]{algorithm2e}
\usepackage[T1]{fontenc}
\usepackage[utf8]{inputenc}

\usepackage{microtype}

\usepackage{inconsolata}

\usepackage[normalem]{ulem}
\usepackage{multirow}
\usepackage{enumitem}

\usepackage{booktabs}

\usepackage[most]{tcolorbox}

\title{Multi-Source Retrieval and Reasoning for Legal Sentencing Prediction}

\usepackage{authblk}

\author[1,2]{\textbf{Junjie Chen}\thanks{chenjj826@gmail.com}}
\author[1]{\textbf{Haitao Li}}
\author[2]{\textbf{Qilei Zhang}}
\author[1]{\textbf{Zhenghua Li}}
\author[3]{\textbf{Ya Zhang}}
\author[3]{\textbf{Quan Zhou}}
\author[3]{\\ \textbf{Cheng Luo}}
\author[1]{\textbf{Yiqun Liu}}
\author[1]{\textbf{Min Zhang}}
\author[2,4]{\textbf{Yueyue Wu}\thanks{Corresponding Author: wuyueyue@tsinghua.edu.cn}}
\author[2]{\textbf{Dongsheng Guo}}
\author[2,1]{\textbf{Qingyao Ai}\thanks{Corresponding Author: aiqy@tsinghua.edu.cn}}

\affil[1]{Department of Computer Science and Technology, Tsinghua University}
\affil[2]{Quancheng Laboratory}
\affil[3]{MegaTech.AI Inc}
\affil[4]{Tsinghua University Think Tank Center}

\begin{document}
\maketitle

\begin{abstract}
\input{abstract}
\end{abstract}

\section{Introduction}
\input{intro}

\begin{figure*}[t]
    \centering
    \includegraphics[width=\textwidth]{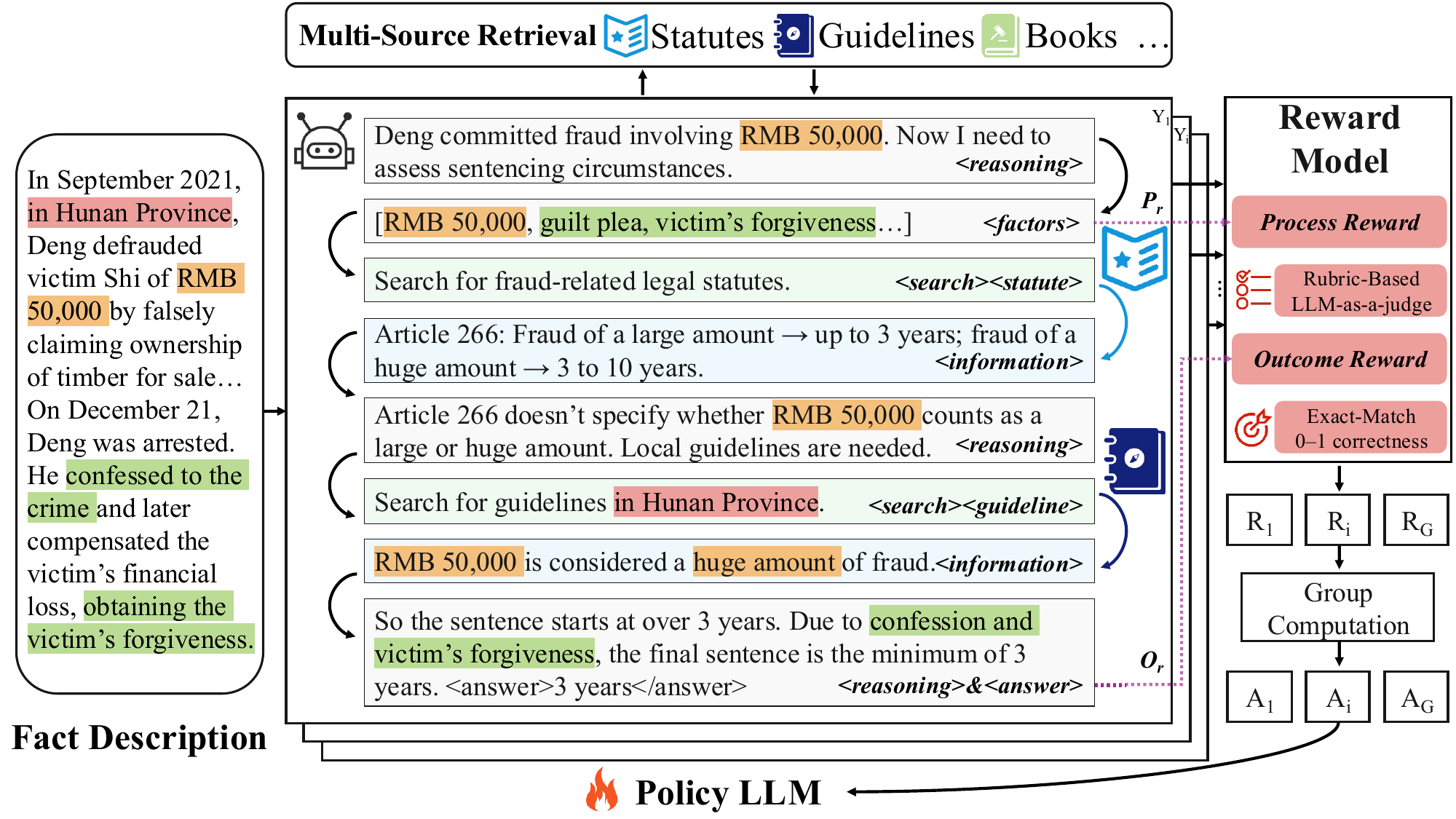}
    \caption{Overview of the $MSR^2$ framework. (1) Multi-source retrieval: the policy LLM issues a <search> request with an optional target source such as <statute> or <guideline>. The query is then routed to the corresponding retriever, and the retrieved top-k evidence is injected into <information> to support further reasoning. (2) Process-level reward: the LLM enumerates both objective and subjective sentencing factors in <factors>. The quality of these factors is evaluated by the LLM-as-a-judge to yield a process-level reward for guidance. (3) RL optimization: the model is trained end-to-end with the GRPO algorithm.}
    \label{fig:1}
\end{figure*}

\section{Related Work}
\textbf{Legal Judgment Prediction.} Legal Judgment Prediction (LJP) aims to predict judicial outcomes, including applicable articles, charges, and sentencing terms, from case facts. Existing methods can be broadly grouped into semantic-focused and logic-focused paradigms. (1) Semantic-focused methods learn representations of case facts for classification or augment them with retrieved precedents~\cite{xu2020distinguish,yue2021neurjudge,zhang2023contrastive,gan2023exploiting,wu2023precedent}. While effective in capturing textual similarity, they often lack the legal rigor required for sentencing and may overlook fine-grained sources such as sentencing guidelines and judicial interpretations. (2) Logic-focused methods improve legal grounding by modeling task dependencies, incorporating logical constraints, or prompting LLMs to follow structured judicial reasoning~\cite{zhong2018legal,gan2021judgment,jiang2023legal,deng2024enabling}. However, they are often limited by rigid formalisms or the intrinsic reasoning capacity of LLMs, making it difficult to handle case-specific nuances and integrate external legal knowledge into sentencing reasoning.

\textbf{RL-Driven Reasoning and Retrieval.} Recent LLM reasoning advances combine inference-time reasoning with training-time optimization~\cite{jaech2024openai,wei2022chain,guo2025deepseek}. Reinforcement learning has further enabled models to develop stronger reasoning behaviors, while search-augmented methods train LLMs to interleave reasoning with retrieval~\cite{jin2025search,zheng2025deepresearcher}. Inspired by these advances, we propose $MSR^2$, an RL-driven framework for legal sentencing prediction. It combines multi-source retrieval with a process-level reward, enabling the model to retrieve relevant legal sources and learn intermediate sentencing-factor identification. This design supports more reliable and interpretable sentencing reasoning. Detailed discussions of related work are provided in Appendix~\ref{app:related_work_details}.

\section{Methodology}
\input{method}

\begin{table}[t]
\centering
\resizebox{\columnwidth}{!}{%
\begin{tabular}{lrr}
\toprule
\textbf{Type} & \textbf{CAIL2018} & \textbf{CJO22} \\
\midrule
Number of Law Articles       & 122   & 66   \\
Number of Charges            & 42    & 42    \\
Number of Prison-Term Classes & 10    & 10    \\
Number of Samples            & 82,138 & 1,698  \\
Average Fact Length (words) & 316.0 & 1,169.6 \\
\bottomrule
\end{tabular}
}
\caption{Statistics of datasets.}
\label{tab:dataset-stats}
\end{table}

\section{Experimental Settings}
\label{exp-setting}
\input{setup}

\section{Results and Analysis}
\input{result}

\section{Conclusion}
We present $MSR^2$, an RL-driven framework that integrates multi-source retrieval with a process-level reward for legal sentencing prediction. By capturing fine-grained objective knowledge and enabling flexible subjective reasoning, $MSR^2$ achieves state-of-the-art performance on CJO22 and CAIL2018 while producing more transparent and traceable reasoning paths, paving the way toward more trustworthy legal AI.

\section{Limitations}
Our study has two main limitations. First, due to resource constraints, we train and evaluate $MSR^2$ mainly with a relatively small backbone, Qwen3-8B. Larger models may further improve both sentencing accuracy and reasoning quality, and we plan to explore this direction in future work. Second, following prior work, we use CAIL2018 and CJO22 as our primary datasets. Although training and testing on CAIL2018 and additionally evaluating on CJO22 provide an initial test of cross-dataset generalization, the evaluation scope remains limited. Future work will extend the evaluation to newer and broader datasets, including more recent cases, regionally diverse legal materials, and cross-jurisdictional or cross-cultural settings.

\section{Ethical Considerations}
Throughout this research, ethical considerations have been integral to ensuring the responsible development and application of AI technologies. We are committed to the principles of open research and the scientific value of reproducibility. Therefore, we have made all the code from our study publicly accessible on GitHub. This transparency allows the community to validate our results and promotes the use of our methods in various settings. Moreover, we used AI assistants only for language polishing and editing to improve clarity, grammar, and readability.

\section*{Acknowledgments}
This work is supported by the Research Project of Quancheng Laboratory, China (Grant No. QCL20250105), and the Key Funding Program of the Tsinghua University Laboratory for Computational Social Science and National Governance (Grant No. CGlab20250002).

% \clearpage 
\bibliography{main}

\appendix
\input{appendix}
\end{document}

%% file: abstract.tex
Legal judgment prediction (LJP) aims to predict judicial outcomes from case facts and typically includes law article, charge, and sentencing prediction. While recent methods perform well on the first two subtasks, legal sentencing prediction (LSP) remains difficult due to its need for fine-grained objective knowledge and flexible subjective reasoning. To address these limitations, we propose $MSR^2$, a framework that integrates multi-source retrieval and reasoning in LLMs with reinforcement learning. $MSR^2$ enables LLMs to perform multi-source retrieval based on reasoning needs and applies a process-level reward to guide intermediate subjective reasoning steps. Experiments on two real-world datasets show that $MSR^2$ improves both accuracy and interpretability in LSP, providing a promising step toward practical legal AI. Our code is available at \url{https://github.com/cjj826/MSR2}. 

%% file: intro.tex
\label{section:intro}
Legal judgment prediction (LJP) aims to predict judicial outcomes based on the textual description of case facts. It is typically formulated as three subtasks: law article prediction, i.e., identifying which specific statutes or legal clauses (articles) apply to a given set of case facts; charge prediction, i.e., determining the formal criminal offense (the name of the crime) the defendant should be charged with; and legal sentencing prediction (LSP), i.e., estimating the punishment (usually the length of a prison term). As a key task in legal AI, LJP has received increasing attention due to its potential to improve judicial efficiency and support the development of practical legal AI systems.

\begin{figure}[t]
    \centering
    \includegraphics[width=\columnwidth]{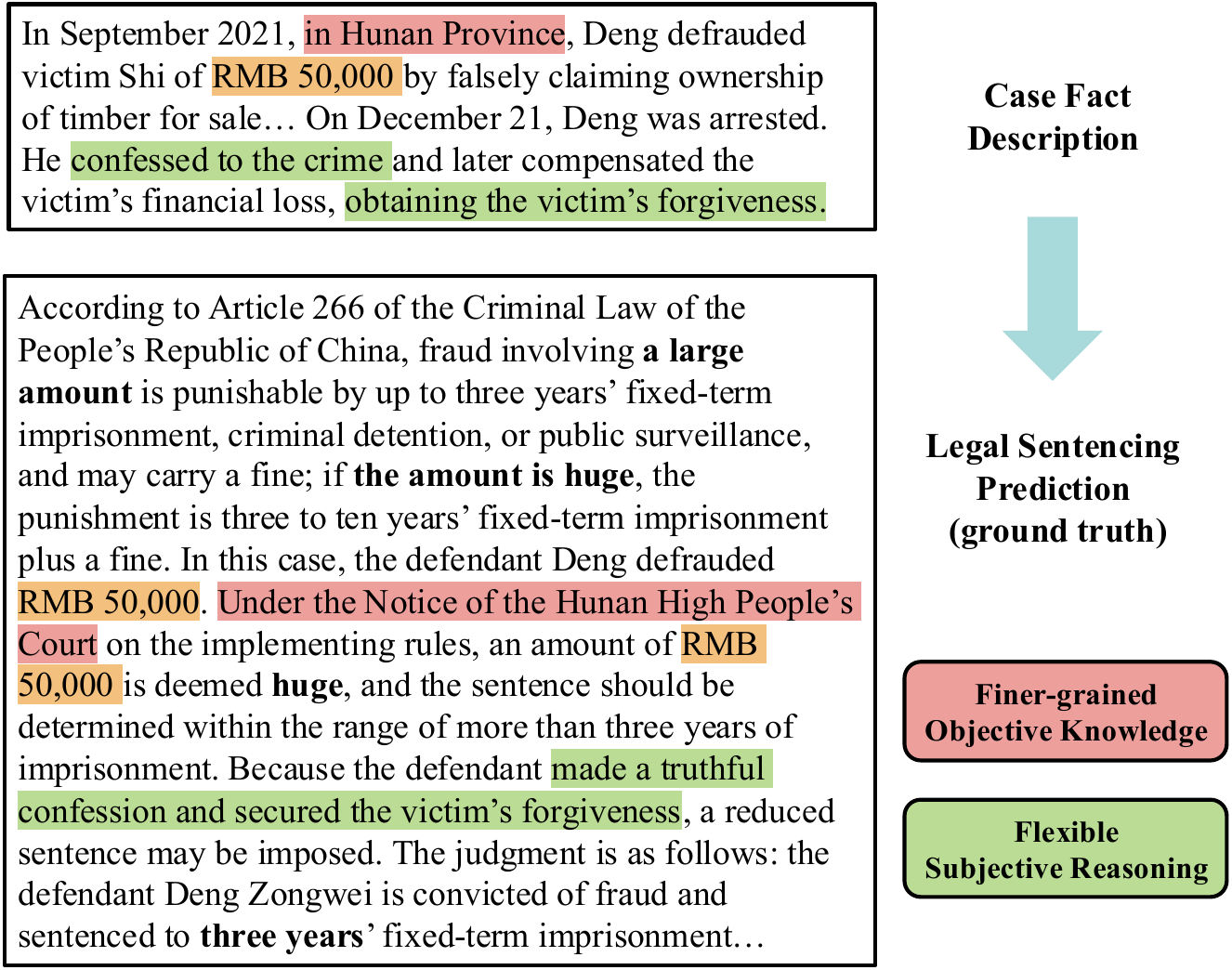}
    \caption{A concrete example of legal sentencing prediction. The case facts at the top are used as input. The bottom shows the human judge’s reasoning and final sentence. This example shows that sentencing prediction needs finer-grained objective knowledge (e.g., \textit{large} / \textit{huge} / \textit{especially huge} thresholds) and flexible subjective reasoning (e.g., mitigation). Existing methods often miss these elements, while our method fills this gap.}
    \label{intro}
\end{figure}

Existing methods for LJP can be broadly grouped into two paradigms. The first is semantic-focused methods, which either learn informative representations of case facts for classification~\cite{xu2020distinguish,xu2024distinguish,yue2021neurjudge,dong2021legal, qin2024explicitly,yue2024circumstance} or incorporate relevant precedents as external information to enrich the context and better model the semantics of the case documents~\cite{wu2023precedent}. The second is logic-focused methods, which strengthen prediction by modeling structural dependencies among subtasks~\cite{zhong2018legal}, injecting legal knowledge through first-order predicate logic constraints~\cite{gan2021judgment}, or prompting large language models (LLMs)~\cite{zhao2026survey} to follow legal syllogism relying solely on their intrinsic reasoning ability~\cite{deng2023syllogistic,jiang2023legal}. 
These methods have been shown to be highly effective in predicting the law article and charge of different legal cases, and have already been used in legal practice as supplementary tools to help judges make decisions. 
As for sentencing prediction, however, it is still considered to be a difficult task for legal AI, and none of the existing methods can provide satisfactory results yet.   

To better understand this phenomenon, we conduct case studies (e.g., Figure~\ref{intro}) and find that sentencing prediction is more complicated than other LJP tasks from both objective and subjective perspectives. First, from the objective perspective,  existing methods often fail to gather important fine-grained information from different legal resources that are necessary for an accurate sentencing prediction. Sentencing often depends on detailed regulatory criteria, such as thresholds that differentiate \textit{large}, \textit{huge}, and \textit{especially huge} amounts, or conditions used to define \textit{especially egregious circumstances}\footnote{We use italics for legal terminology to differentiate it from regular text.}. These standards are rarely stated in statutes, but are scattered across judicial interpretations, sentencing guidelines, and other legal documents. Current approaches primarily utilize statutes and precedents, failing to incorporate this fragmented yet crucial body of knowledge, which limits their ability to support accurate sentencing.

Second, from the subjective perspective, existing methods often lack sufficient capability for the subjective reasoning required for LSP. Sentencing often involves judgments about subjective (or discretionary) sentencing factors, such as the defendant's mental state (e.g., intent or negligence), or determinations of whether a theft constitutes \textit{pickpocketing}, which requires evaluating whether the act occurred in \textit{a public place} and whether the property was \textit{carried on the person}. Such subjective determinations involve inherently ambiguous concepts and typically require deeply synthesizing information from multiple sources together with case-specific context to make reasonable inferences. Approaches that only model superficial topological structures, rely on rigid formalisms such as first-order predicate logic with limited expressiveness, or depend solely on the intrinsic capabilities of LLMs, are not flexible and strong enough to conduct subjective reasoning for legal sentencing prediction.

Based on these observations, we argue that an effective sentencing method should (i) flexibly retrieve external knowledge from the appropriate sources based on reasoning needs, and (ii) tightly integrate the retrieved legal information into the reasoning to support broader and deeper inference for legal sentencing prediction.

% Recent advances such as OpenAI’s o-series~\cite{jaech2024openai} and DeepSeek-R1~\cite{guo2025deepseek} demonstrate that large language models (LLMs) trained with reinforcement learning with verifiable rewards (RLVR) can exhibit remarkable capabilities in both reasoning and coding. At the same time, efforts like Search-R1 have advanced the integration of retrieval and reasoning in LLMs~\cite{jin2025search}. Inspired by these insights
Inspired by recent advances in reinforcement learning~\cite{jaech2024openai,guo2025deepseek} and reasoning-based retrieval methods~\cite{jin2025search}, we propose $MSR^2$, a framework that integrates multi-source retrieval with reasoning for LSP using RL. Specifically, $MSR^2$ addresses both the objective and subjective challenges from two directions: 
(1) We introduce a multi-source retrieval mechanism that allows the LLM to query appropriate legal sources, such as statutes, judicial interpretations, and sentencing guidelines, based on its reasoning needs and to continue reasoning with the retrieved information. 
(2) We introduce a process-level reward on the reasoning process of the sentencing prediction models in addition to the traditional outcome reward widely used in reinforcement learning~\cite{team2025kimi}. This mechanism guides the LLM’s intermediate reasoning steps, enhancing its capability to synthesize retrieved information with case facts for accurate discretionary determinations, thereby achieving a tighter integration of retrieval and reasoning. 
Through these methods, we allow $MSR^2$ to dynamically incorporate fine-grained and diverse external knowledge into its subjective reasoning process to produce explainable and reliable sentencing.
Experiments on two real-world datasets, CAIL2018~\cite{xiao2018cail2018} and CJO22~\cite{wu2023precedent},  show that $MSR^2$ significantly improves predictive accuracy and interpretability, providing a promising path toward more trustworthy and practical legal AI.

%% file: method.tex
\subsection{Task Definition}

In this paper, we focus on \textbf{Legal Sentencing Prediction (LSP)}. 
Most existing methods formulate LSP as multi-class classification 
over predefined sentencing intervals (e.g., $\leq 6$ months, 6--9 months, 
9--12 months, $\dots$, $> 10$ years). However, such 
interval-based discretization may overlook fine-grained differences between 
cases within the same category. To address this, we consider a more fine-grained setting: predicting the sentence value before mapping it to the standard sentencing interval.

Formally, given a fact description $x$ (Fig.~\ref{fig:1}, left), the model outputs a sentence value $\hat{y}$, which is mapped to the corresponding interval for fair comparison with prior work. This setting is more challenging 
than direct interval classification because it demands fine-grained reasoning, yet it 
better reflects real-world sentencing practice. Details of the evaluation 
settings are provided in Section~\ref{exp-setting}.

\subsection{Method Overview}
As discussed in Section~\ref{section:intro}, an effective sentencing model should not only retrieve relevant legal information but also tightly integrate it into the reasoning. To this end, we propose $MSR^2$, a framework with two key components: a multi-source retrieval mechanism and a process-level reward.

Figure~\ref{fig:1} (middle) illustrates one rollout of the policy LLM under our framework. The LLM follows the official sentencing logic proposed by the Supreme People’s Court\footnote{https://www.lawinfochina.com}. First, the model reads the fact description $x$ and writes its initial analysis in <reasoning>. It then handles the most difficult step, identifying sentencing factors, by extracting both objective and subjective factors and recording them in <factors>. When a knowledge gap appears, the model sends a <search> query, which is routed to an appropriate legal source; the retrieved text is returned in <information> and used to continue reasoning. Finally, the model outputs a determinate sentence in <answer>. Next, we describe the multi-source retrieval, the process-level reward, and the training with Group Relative Policy Optimization (GRPO)~\cite{shao2024deepseekmath}. 
Algorithm~\ref{alg:rollout-router} in Appendix~\ref{Rollout Algorithm} shows the LLM rollout with multi-source retrieval.

\subsection{Multi-Source Retrieval}
\label{sec:retrieval}
We define a set of legal sources $S=\{S_1, S_2, \ldots, S_n\}$, where each $S_i$ denotes a source type such as statutes, sentencing guidelines, or judicial interpretations. During reasoning, when required information is missing, the policy LLM issues a request in <search>. When a <search> tag appears, we parse its content to form a query $q_t$ and route the query to an appropriate information source $s_t$. A key challenge is how to perform this routing from $q_t$ to $s_t$. To address this, we consider several routing strategies: (1) Action-space routing. We can treat routing as part of the policy action. In this routing strategy, the policy LLM outputs both the query content and a target source tag, such as <statute> when a law article is needed or <guideline> when an amount threshold must be checked, and the predicted tag directly determines $s_t$. This makes routing an explicit decision in the reasoning trajectory and allows it to be optimized jointly with the policy during training. (2) MoE-style dispatcher. We can view each retriever $\mathcal{R}_{S_i}$ as an expert and learn a lightweight dispatcher. Given a representation of the query $q_t$, a linear gating layer produces a distribution over sources, and we select $s_t$ via top-1 (or top-$m$) gating. (3) Dedicated Routing Agent. We can employ a dedicated agent that takes the reasoning context and query as input and selects a suitable source $s_t$.

In this work, we adopt the first strategy and treat routing as part of the policy LLM's action, as this design is simple and introduces no additional overhead. Our framework can be readily extended to other routing strategies, and a more systematic comparison of routing strategies together with a fine-grained analysis of routing behavior is an interesting direction for future work. Once routed to the matching $\mathcal{R}_{s_t}$, the retriever returns the top-$k$ items:
\begin{equation}
E_t^{(s_t)}=\mathrm{TopK}\!\big(\mathcal{R}_{s_t}(q_t),\,k\big)
=\{e_{t,1},\ldots,e_{t,k}\}.
\label{eq:topk}
\end{equation}
The retrieved items are placed in <information> and appended to the reasoning trajectory to continue inference. This multi-source retrieval enables the policy LLM to access the appropriate source when needed, and allows flexible retrieval strategies for different sources.

\subsection{Process-Level Reward}
Relying solely on the outcome reward can fail to provide meaningful feedback when the final prediction is correct, but the reasoning is wrong. In such cases, the model may not enhance the subjective reasoning ability that accurate sentencing prediction requires. To strengthen supervision, we introduce a process-level reward that focuses on the model’s performance in the most challenging step: identifying sentencing factors. These factors include both objective factors (e.g., amount at stake and number of offenses) and subjective factors (e.g., whether the act occurred in a public place, whether property was carried on the person, and the defendant’s attitude toward confession).

Let the extracted factor set be $F=\{f_1,\ldots,f_m\}$. We then evaluate its quality and obtain a score. The scoring method is flexible: it can be rule-based, human-annotated, or produced by the LLM-as-a-judge~\cite{li2024llms,gu2026survey}. Our framework does not depend on a specific choice and can adapt to different settings. In our experiments, we use the LLM-as-a-judge; implementation details are provided in  Section~\ref{Implementation Details}.

Given the score, we normalize it to the range $[0,1]$ to obtain the process reward $P_r$. We further compute an outcome reward $O_r$ for the final prediction. To stay consistent with prior work~\cite{zhong2018legal,wu2023precedent}, we map both the ground-truth sentence (a concrete value) and our model prediction to a discrete interval and use classification accuracy as the outcome score, i.e., $O_r=1$ if the predicted interval is correct and $O_r=0$ otherwise. The total reward is the weighted sum of the process reward and the outcome reward:

\begin{equation}
R = (1-\lambda_r) O_r + \lambda_r P_r.
\label{eq:R-total}
\end{equation}
Here, $\lambda_r$ is the hyperparameter.

\subsection{Training with GRPO}
\label{sec:grpo}

We train the policy LLM $\pi_\theta$ with Group Relative Policy Optimization (GRPO)~\cite{shao2024deepseekmath}. Details are in Appendix~\ref{GRPO Training Details}.

%% file: setup.tex
% \begin{table}[t]
% \centering
% \caption{Statistics of datasets.}
% \label{tab:dataset-stats}
% \resizebox{\columnwidth}{!}{%
% \begin{tabular}{lrr}
% \toprule
% \textbf{Type} & \textbf{CAIL2018} & \textbf{CJO22} \\
% \midrule
% \# Law Article       & 122   & 66   \\
% \# Charge            & 42    & 42    \\
% \# Prison Term       & 10    & 10    \\
% \# Sample            & 82138 & 1698  \\
% Avg.\ \# words in Fact & 316.0 & 1,169.6 \\
% \bottomrule
% \end{tabular}
% }
% \end{table}

\subsection{Datasets and Evaluation Metrics}
We follow prior LJP studies~\cite{zhong2018legal,xu2020distinguish,xu2024distinguish,yue2021neurjudge,wu2023precedent} and conduct experiments on CAIL2018 and CJO22~\cite{wu2023precedent}. 
For CAIL2018, we follow the preprocessing protocol of PLJP~\cite{wu2023precedent}. Specifically, we use the \textit{exercise} and \textit{final} subsets, remove cases with the death penalty, life imprisonment, or zero term of imprisonment, and further filter out cases involving multiple charges or multiple applicable law articles. 
After filtering, we randomly sample 82{,}138 cases and split them into training, validation, and test sets with an 8:1:1 ratio. 
For CJO22, we use it only as an additional test set following PLJP~\cite{wu2023precedent}. 
Dataset statistics are reported in Table~\ref{tab:dataset-stats}. For evaluation, we report Accuracy, Macro-Precision (Ma-P), Macro-Recall (Ma-R), and Macro-F1 (Ma-F), following prior work~\cite{zhong2018legal,xu2020distinguish,xu2024distinguish,yue2021neurjudge,wu2023precedent}.

\subsection{Baselines}
We compare $MSR^2$ against various baselines, including conventional classifiers, neural LJP models, and recent LLM-based methods.

\textbf{Classification baselines.} We include \textbf{TFIDF+SVM}~\cite{cortes1995support}, a sparse lexical feature baseline with a linear classifier; \textbf{CNN}~\cite{lecun1989backpropagation}, which applies convolutional kernels to capture local n-gram features for classification; \textbf{BERT}~\cite{devlin2019bert}, a pre-trained encoder fine-tuned for LJP; and \textbf{RoBERTa}~\cite{liu2019roberta}, a robustly optimized BERT variant fine-tuned for LJP. We also include representative LJP models that explicitly model task structure: \textbf{TopJudge}~\cite{zhong2018legal}, a multi-task framework that captures dependencies among LJP subtasks; \textbf{LADAN}~\cite{xu2020distinguish}, which leverages law-article relations and distillation to learn discriminative representations for confusable articles; and \textbf{NeurJudge}~\cite{yue2021neurjudge}, which decomposes fact descriptions into circumstance-aware parts for prediction.

\textbf{LLM-based baselines.} We evaluate both direct inference and direct reasoning. We evaluate Llama3-8B~\cite{grattafiori2024llama} and the Qwen3 series (8B, 14B, 32B, 235B-A22B)~\cite{yang2025qwen3} using a prompting protocol modeled after the legal syllogism and real-world judicial sentencing steps. Direct inference produces the final prediction directly, while direct reasoning prompts the model to generate explicit intermediate reasoning before outputting the prediction.

\textbf{Retrieval-augmented generation.} We include \textbf{PLJP}~\cite{wu2023precedent} (implemented with GPT-4o-mini) and two RAG variants based on Qwen3-8B and Qwen3-14B. All variants use the identical retrieval modules of our method.

\textbf{Training-based methods.} We fine-tune Llama3-8B and Qwen3-8B on the training set using LLaMA-Factory~\cite{zheng2024llamafactory} with LoRA~\cite{hu2021lora}  for one epoch.

\begin{table*}[t]
\centering
\resizebox{0.9\textwidth}{!}{%
\begin{tabular}{lrrrrrrrr}
\hline
\textbf{Method}     & \multicolumn{4}{c}{\textbf{CJO22}}                                & \multicolumn{4}{c}{\textbf{CAIL2018}}                             \\ \cline{2-9} 
\textbf{}           & \textbf{Acc}   & \textbf{Ma-P}  & \textbf{Ma-R}  & \textbf{Ma-F}  & \textbf{Acc}   & \textbf{Ma-P}  & \textbf{Ma-R}  & \textbf{Ma-F}  \\ \hline
\multicolumn{9}{l}{\textit{\textbf{Classification methods}}}                                                                                                \\
TFIDF+SVM~\cite{cortes1995support}           & 33.16          & 23.17          & 19.70          & 20.01          & 42.05          & 28.56          & 26.42          & 26.66          \\
CNN~\cite{lecun1989backpropagation}                 & 34.28          & 15.34          & 16.47          & 13.79          & 39.39          & 19.38          & 17.51          & 16.08          \\
BERT~\cite{devlin2019bert}                & 33.57          & 19.86          & 17.88          & 16.47          & 42.34          & 23.78          & 23.51          & 22.64          \\
RoBERTa~\cite{liu2019roberta}             & 35.34          & 23.11          & 20.24          & 19.27          & 42.12          & 28.90          & 24.14          & 23.08          \\
TopJudge~\cite{zhong2018legal}            & 27.14          & 19.76          & 17.69          & 17.94          & 35.70          & 18.59          & 18.43          & 17.63             \\
LADAN~\cite{xu2020distinguish}               & 33.69          & 26.40          & 22.94          & 24.55          & 36.14          & 20.86          & 18.03          &  16.58              \\
NeurJudge~\cite{yue2021neurjudge}           & 26.80          & 26.81          & 26.85          & 25.97          & 36.84          & 20.99          & 16.81          &  18.51              \\ \hline
\multicolumn{9}{l}{\textit{\textbf{Direct inference}}}                                                                                                      \\
Llama3-8B-DI           & 12.96          & 16.63          & 12.45          & 8.69           & 12.54          & 14.91          & 13.12          & 8.60           \\
Qwen3-8B-DI            & 15.14          & 35.81          & 15.40          & 11.28          & 11.38          & 31.92          & 14.04          & 8.72           \\
Qwen3-14B-DI           & 14.43 & 36.78 & 18.14 & 12.23 & 12.82 & 26.41 & 16.35 & 10.92 \\
Qwen3-32B-DI           & 28.45          & 27.82          & 22.67          & 22.06          & 26.49          & 27.68          & 27.28          & 23.43          \\
Qwen3-235B-A22B-DI     & 37.28          & 34.81          & 26.35          & 27.29          & 38.05          & 29.36          & 28.55          & 26.82          \\ \hline
\multicolumn{9}{l}{\textit{\textbf{Direct reasoning}}}                                                                                                      \\
Qwen3-8B-DR            & 22.67          & 22.91          & 22.87          & 18.87          & 20.57          & 20.43          & 21.07          & 16.73          \\
Qwen3-14B-DR           & 27.80          & 24.18          & 26.51          & 22.42          & 25.58          & 19.35          & 22.44          & 18.03          \\
Qwen3-32B-DR           & 30.09          & 26.20          & 27.96          & 24.53          &   27.69          &         23.21	      &       26.57	        &  22.22               \\ \hline
\multicolumn{9}{l}{\textit{\textbf{Retrieval-augmented generation}}}                                                                                         \\
PLJP-GPT-4o-mini~\cite{wu2023precedent}                &  27.92         & 24.88          &  20.90         &  18.63         &  27.53         & 23.65          &  19.56         &  18.03          \\
Qwen3-8B-RAG            & 29.03          & 22.75          & 24.26          & 20.95          & 28.16          & 20.54          & 21.56          & 18.70          \\
Qwen3-14B-RAG           & 30.57          & 27.07          & 24.05          & 23.81          & 29.68          & 22.82          & 21.00          & 19.57          \\ \hline
\multicolumn{9}{l}{\textit{\textbf{Fine-tuning}}}                                                                                                                   \\
Llama3-8B-FT           & 31.68          & 32.05          & 19.48          & 17.26          & 37.15          & 26.83          & 21.37          & 19.38          \\
Qwen3-8B-FT            & 37.63          & 36.45          & 21.85          & 19.87          & 41.84          & 29.57          & 23.89          & 22.31          \\ \hline
$MSR^2$-Qwen3-8B (ours) & \textbf{44.29} & \textbf{37.43} & \textbf{30.73} & \textbf{28.07} & \textbf{46.18} & \textbf{36.74} & \textbf{31.70} & \textbf{28.84} \\
\bottomrule
\end{tabular}
}
\caption{Results on CJO22 and CAIL2018. We report Accuracy (Acc), Macro-Precision (Ma-P), Macro-Recall (Ma-R), and Macro-F1 (Ma-F). The best numbers are in bold; $MSR^2$ achieves the top accuracy across datasets.}
\label{tab:my-table}
\end{table*}

\subsection{Implementation Details}
\label{Implementation Details}

\subsubsection{Training}
We use Qwen3-8B as the backbone policy model and train it with verl. The full training configuration is provided in Appendix~\ref{Training Details}.
\subsubsection{Retrieval}
We construct four retrieval sources, including statutes, legal books, precedents, and sentencing guidelines. Each source is served by a dedicated retriever under a unified interface, and we return the top-$k$ results with $k=3$ by default. To prevent evaluation leakage, we carefully inspect all retrieval sources before experiments.

\textbf{Statutes.} The corpus contains 51{,}784 statutes. We encode each statute into a 1{,}024-dimensional vector using Qwen3-Embedding and perform dense retrieval over the full text with FAISS, using an HNSW index and cosine similarity.

\textbf{Legal books.} The legal book corpus contains around 1{,}200 books. We segment them into 780{,}035 passages and encode each passage into a 1{,}024-dimensional vector using Qwen3-Embedding. At query time, we perform hybrid retrieval by combining BM25 with dense retrieval based on an HNSW index and cosine similarity over the passage text.

\textbf{Precedents.} The precedent corpus contains 10{,}714{,}118 criminal judgment documents. We use BM25 over the full text, which is effective for matching offense names, numeric cues, and template-like legal phrasing.

\textbf{Sentencing guidelines.} The sentencing guideline corpus consists of 200 official guideline documents released by different provinces and municipalities, covering a wide range of common offenses. We also use BM25 over the full text for this source.

\subsubsection{Process-level Reward}

We employ Qwen3-32B as the LLM-as-a-judge and set $\lambda_r=0.2$. To standardize the evaluation, the judge is guided by a rubric-based instruction (illustrated in Figure~\ref{fig:prompt_scoring} in Appendix~\ref{Process-level Reward Details}), which specifies concrete criteria to assess whether each factor is empirically supported by the case facts. This design follows recent work on using rubrics as structured reward signals, which provide more stable and controllable supervision than direct scalar ratings~\cite{gunjal2026rubrics}.

For the hyperparameters involved in training, retrieval, and process-level reward, we choose their values based on preliminary validation and computational considerations. We further analyze several key choices that are central to our framework. For example, in Section~\ref{ablation_study_section}, we compare the full model ($\lambda_r=0.2$) with a variant that removes the process-level reward ($\lambda_r=0$). In Section~\ref{Reliability of LLM-as-a-Judge}, we also examine the effect of different LLM-as-a-judge models. However, a comprehensive sensitivity study over all hyperparameters would require substantial additional training runs. We therefore focus on validating the main design choices in this paper, while a finer-grained analysis of these hyperparameters is an interesting direction for future work.

%% file: result.tex
\begin{table}[t]
\centering
\resizebox{0.9\columnwidth}{!}{%
\begin{tabular}{lrrrr}
\hline
\textbf{CJO22} & \textbf{Acc} & \textbf{Ma-P} & \textbf{Ma-R} & \textbf{Ma-F} \\ \hline
\multicolumn{5}{l}{\textit{training step 0}}                                                    \\
$MSR^2$            & 29.03        & 22.75         & 24.26         & 20.95         \\
w/o PR         & 29.03        & 22.75         & 24.26         & 20.95         \\
w/o MR         & 28.09        & 24.50         & 22.94         & 21.77         \\ \hline
\multicolumn{5}{l}{\textit{training step 40}}                                                   \\
$MSR^2$            & 35.92        & 32.61         & 28.12         & 25.07         \\
w/o PR         & 33.04        & 24.59         & 23.11         & 22.47         \\
w/o MR         & 32.69        & 23.92         & 22.80          & 21.85         \\ \hline
\multicolumn{5}{l}{\textit{training step 80}}                                                   \\
$MSR^2$            & 35.57        & 26.16         & 29.08         & 26.12         \\
w/o PR         & 34.51        & 25.13         & 23.78         & 23.33         \\
w/o MR         & 34.39        & 24.87         & 24.78         & 23.14         \\ \hline
\multicolumn{5}{l}{\textit{training step 100}}                                                  \\
$MSR^2$            & 37.40        & 32.12         & 29.26         & 26.02         \\
w/o PR         & 33.51        & 25.33         & 25.46         & 23.59         \\
w/o MR         & 35.10        & 26.50         & 25.49         & 24.07         \\ \hline
\multicolumn{5}{l}{\textit{training step 120}}                                                  \\
$MSR^2$            & 38.40        & 30.54         & 28.10         & 24.96         \\
w/o PR         & 35.51        & 25.15         & 25.12         & 23.25         \\
w/o MR         & 34.22        & 25.17         & 24.68         & 22.96         \\ \hline
\multicolumn{5}{l}{\textit{training step 140}}                                                  \\
$MSR^2$            & 38.40        & 30.24         & 28.31         & 24.44         \\
w/o PR         & 36.22        & 25.70         & 24.31         & 22.62         \\
w/o MR         & 35.10        & 27.44         & 22.89         & 22.90         \\ \hline
\end{tabular}
}
\caption{Effect of the process reward (PR) and multi-source retrieval (MR) on CJO22.}
\label{tab:pr-cjo22}
\end{table}

In this section, we present the experimental results and aim to address the following four research questions (RQs):

\noindent\textbf{RQ1: } How does $MSR^2$ perform compared to other baselines?

\noindent\textbf{RQ2: } How does each component of $MSR^2$ contribute to the final performance?

\noindent\textbf{RQ3: } Can $MSR^2$ provide interpretable reasoning for cases?

\noindent\textbf{RQ4: } How reliable are the process-level rewards obtained through LLM-as-a-judge?

\subsection{Main Results (RQ1)}

To validate the effectiveness of $MSR^2$, we compare it with a broad set of baselines. Table~\ref{tab:my-table} summarizes the overall results on CAIL2018 and CJO22. Based on the results, we can draw the following conclusions: 

\textbf{$MSR^2$ achieves the best overall performance.}
Across both CAIL2018 and CJO22, $MSR^2$ consistently ranks at the top and improves not only accuracy but also macro metrics, indicating more balanced gains across categories.

\textbf{Conventional LJP models are more sensitive to distribution shift.}
While traditional classifiers and earlier neural models can remain competitive on CAIL2018, their performance degrades more clearly on CJO22. This trend suggests that methods relying heavily on dataset-specific textual semantic similarity generalize less reliably.

\textbf{For LLMs, direct reasoning is consistently stronger than direct inference, but still insufficient.}
Prompting LLMs to produce explicit intermediate reasoning generally outperforms direct inference, showing that making the reasoning process explicit helps. However, the gap to $MSR^2$ remains, implying that sentencing prediction requires more than prompting LLMs to reason internally.

\textbf{Retrieval augmentation helps, but naive retrieval integration leaves substantial room for improvement.}
Retrieval-augmented baselines improve over prompting-only LLM settings, consistent with the motivation that external information can support judgment prediction. Yet, these baselines still fall behind $MSR^2$, highlighting that the key is not merely adding retrieval, but integrating retrieved information into the reasoning.

\textbf{Supervised fine-tuning provides strong gains, but $MSR^2$ further improves beyond it.}
Fine-tuning substantially strengthens LLM baselines compared to prompting-only settings, confirming the value of task-specific adaptation. $MSR^2$ remains superior, suggesting that reinforcement learning with multi-source retrieval and process-level reward provides additional benefits beyond standard fine-tuning.

Interestingly, our $MSR^2$ initialized from Qwen3-8B substantially outperforms much larger Qwen3 variants under direct prompting (e.g., Qwen3-235B-A22B) and direct reasoning (e.g., Qwen3-32B). This suggests that the gains from our method can exceed those obtained by scaling up the backbone model alone. As a result, $MSR^2$ offers a more cost-effective path and provides practical insights for deployment.

\subsection{Ablation Study (RQ2)}
\label{ablation_study_section}

To quantify the individual contribution of each component, we conduct an ablation study on the CJO22 dataset by evaluating $MSR^2$ against two specific variants: (1) w/o PR, which removes the process-level reward while retaining only the outcome reward, and (2) w/o MR, which disables multi-source retrieval and relies exclusively on statutory article retrieval. Table~\ref{tab:pr-cjo22} reports the comparative results across different training steps.

\textbf{Overall, $MSR^2$ consistently achieves the best performance across checkpoints, indicating that PR and MR are complementary rather than interchangeable.} At training step 0, $MSR^2$ and w/o PR are identical, which is expected since the benefit of PR emerges through reinforcement learning updates. As training proceeds, removing either component yields a clear drop. For example, at training step 40, $MSR^2$ reaches 35.92 Acc and 25.07 Ma-F, while w/o PR drops to 33.04 Acc and 22.47 Ma-F, and w/o MR further drops to 32.69 Acc and 21.85 Ma-F. 

\textbf{MR becomes increasingly important as optimization progresses, especially for final accuracy.} At training step 120, $MSR^2$ achieves 38.40 Acc, whereas w/o PR reaches 35.51 Acc and w/o MR only 34.22 Acc, suggesting that accurate retrieval from the right sources is critical for converging to stronger policies. 

Overall, PR improves intermediate factor identification, MR provides external information needed for reliable sentencing prediction, and their combination yields the most robust gains.

\subsection{Case Analysis (RQ3)}
In this section, we present a fraud case study to illustrate the accuracy and interpretability of $MSR^2$. As shown in Figure~\ref{fig:3} in Appendix~\ref{Case Study Details}, $MSR^2$ correctly predicts the sentence of three years and provides a traceable reasoning path from case facts to judgment. It identifies key sentencing factors, including the amount involved, confession, and victim forgiveness, and retrieves both the relevant fraud statute and the Hunan sentencing guidelines. These sources verify that RMB~50{,}000 constitutes a huge amount under local practice, placing the baseline sentence in the three-to-ten-year range; the mitigating factors then support reducing the sentence to the statutory minimum.

In contrast, classification models such as BERT only output coarse sentence intervals without justification. Retrieval-based methods such as PLJP rely mainly on statutes and precedents. Although PLJP cites Article 266, it overlooks fine-grained local guidance and fails to determine whether RMB~50{,}000 constitutes a huge amount under Hunan practice, leading to a lighter recommendation of one to two years. It also makes limited use of subjective mitigating factors such as confession and victim forgiveness. This case suggests that $MSR^2$ can be more reliable because it grounds its prediction in fine-grained legal sources and makes each step of the sentencing adjustment explicit.

\subsection{Reliability of LLM-as-a-Judge (RQ4)}
\label{Reliability of LLM-as-a-Judge}

\begin{table}[t]
\resizebox{\columnwidth}{!}{%
\begin{tabular}{lccc}
\hline
\textbf{Judge Model} & \textbf{Pearson ($r$)}            & \textbf{Spearman ($\rho$)} & \textbf{Kendall ($\tau$)} \\ \hline
Among Humans         & 0.637                    & 0.643                      & 0.590                     \\ \hline
Qwen3-32B            & 0.587$^{\dagger\dagger}$ & 0.554$^{\dagger\dagger}$   & 0.512$^{\dagger\dagger}$  \\
DeepSeek-V3-0324     & 0.451$^{\dagger\dagger}$ & 0.431$^{\dagger\dagger}$   & 0.393$^{\dagger\dagger}$  \\
GPT-4o               & 0.324$^{\dagger\dagger}$ & 0.318$^{\dagger\dagger}$   & 0.297$^{\dagger\dagger}$  \\ \hline
\end{tabular}%
}
\caption{Consistency analysis of LLM-as-a-judge on 200 randomly sampled cases. $^{\dagger\dagger}$ indicates $p$ < 0.01.}
\label{tab:correlation_results}
\end{table}

To assess the reliability of the process-level rewards derived from LLM-as-a-judge, we conduct a human--model agreement study on 200 randomly sampled cases, with annotation details provided in Appendix~\ref{app:expert_annotation}. For each case, sentencing factors are first extracted by our policy model (Qwen3-8B) and then independently rated by three legal experts from leading law schools on a 1--10 scale. We aggregate expert ratings by majority voting when a majority exists, and otherwise use the median score as the final human annotation. Under the same scoring instructions in Figure~\ref{fig:prompt_scoring}, we use Qwen3-32B, DeepSeek-V3-0324, and GPT-4o to rate the same cases, and measure their agreement with the aggregated human scores using Pearson's $r$, Spearman's $\rho$, and Kendall's $\tau$. As shown in Table~\ref{tab:correlation_results}, Qwen3-32B achieves the highest agreement, with correlations closest to the human--human agreement. GPT-4o obtains lower correlations, possibly because it has less exposure to Chinese legal corpora during pretraining, which limits its effectiveness in this specific domain. Nevertheless, all model--human correlations are statistically significant ($p < 0.01$). These results indicate that the process-level rewards produced by LLM-as-a-judge are reasonably aligned with expert judgments. Considering both agreement with human annotations and computational efficiency, we adopt Qwen3-32B as the judge model in our experiments.

%% file: appendix.tex
\section{Related Work Details}
\input{relate_work}

\section{Rollout Algorithm}
\label{Rollout Algorithm}
In this section, we present the pseudocode of LLM rollout with multi-source retrieval in Algorithm~\ref{alg:rollout-router}.
\begin{algorithm}[t]
\caption{LLM Rollout with Multi-Source Retrieval}
\label{alg:rollout-router}
\DontPrintSemicolon
\KwIn{$x$, policy $\pi_\theta$, sources $S=\{s_1,\ldots,s_n\}$, retrievers $\{\mathcal{R}_s\}_{s\in S}$, budget $B$, top-$k$ $k$}
\KwOut{final response $y$}
$y\leftarrow\varnothing$; $b\leftarrow0$ \;
\While{$b<B$}{
  $y_b\leftarrow\varnothing$ \;
  \While{true}{
    sample $y_t\sim\pi_\theta(\cdot\mid x,\,y{+}y_b)$; $y_b\leftarrow y_b{+}y_t$ \;
    \If{$y_t\in\{\langle/search\rangle,\,\langle/answer\rangle,\,\langle eos\rangle\}$}{\textbf{break}}
  }
  $y\leftarrow y{+}y_b$ \;
  \If{\texttt{<search>...</search>} in $y_b$}{
    $(q_t,s_t)\leftarrow\textsc{ParseSearchAndRoute}(y_b)$; $E_t^{(s_t)}\leftarrow\textsc{TopK}(\mathcal{R}_{s_t}(q_t),k)$ \;
    insert $\langle information\rangle E_t^{(s_t)}\langle/information\rangle$ into $y$ \;
  }\ElseIf{\texttt{</answer>} in $y_b$}{\Return $y$}
  \Else{$y\leftarrow y{+}$ ``My action is not correct. Let me rethink.''}
  $b\leftarrow b+1$ \;
}
\Return $y$
\end{algorithm}

\section{GRPO Training Details}
\label{GRPO Training Details}

The Group Relative Policy Optimization (GRPO)~\cite{shao2024deepseekmath} replaces the learned value function with a group-wise baseline computed from multiple sampled trajectories. For each input $x$, we sample a group of $G$ trajectories $Y=\{y_1,\ldots,y_G\}$ from the previous policy $\pi_{\mathrm{old}}$. Let $R_i=R(x,y_i)$, $\bar{R}=\frac{1}{G}\sum_{i=1}^{G}R_i$, and $\sigma_R=\sqrt{\frac{1}{G}\sum_{i=1}^{G}(R_i-\bar{R})^2}$. We define the group-relative advantage for trajectory $y_i$ as
\begin{equation}
\tilde{A}_i \;=\; \frac{R_i-\bar{R}}{\sigma_R+\varepsilon},
\label{eq:adv_group}
\end{equation}
where $\varepsilon$ is a small constant for numerical stability. Since our rollouts interleave model-generated tokens with retrieved tokens, we mask retrieved spans from optimization. Specifically, let $\mathbb{I}(y_{i,t})=1$ if $y_{i,t}$ is generated by the model and $\mathbb{I}(y_{i,t})=0$ if it is a retrieved token, and let $Z_i=\sum_{t=1}^{|y_i|}\mathbb{I}(y_{i,t})$ be the number of generated tokens. The per-token advantage is then
\begin{equation}
\widehat{A}_{i,t} \;=\; \tilde{A}_i \cdot \mathbb{I}(y_{i,t}),
\label{eq:adv}
\end{equation}
so, retrieved tokens do not contribute to the loss.

We define the likelihood ratio for generated tokens:
\begin{equation}
r_{i,t}
\;=\;
\frac{\pi_\theta\!\left(y_{i,t}\mid x, y_{i,<t}; \mathcal{R}\right)}
{\pi_{\mathrm{old}}\!\left(y_{i,t}\mid x, y_{i,<t}; \mathcal{R}\right)} ,
\label{eq:ratio}
\end{equation}
and optimize the GRPO objective: 
\begin{equation}
\resizebox{\columnwidth}{!}{$
\begin{aligned}
\mathcal{J}_{\mathrm{GRPO}}(\theta)
&= \mathbb{E}_{x,\,Y\sim\pi_{\mathrm{old}}(\cdot\mid x;\mathcal{R})}\!\Bigg[
\frac{1}{G}\sum_{i=1}^{G}\frac{1}{Z_i}\sum_{t=1}^{|y_i|}\mathbb{I}(y_{i,t})\,
\min\!\Big( r_{i,t}\,\widehat{A}_{i,t},\,
\mathrm{clip}(r_{i,t},\,1-\epsilon,\,1+\epsilon)\,\widehat{A}_{i,t} \Big)
\Bigg]
\\
&\quad - \beta \, D_{\mathrm{KL}}\!\big[\pi_\theta \,\Vert\, \pi_{\mathrm{ref}}\big],
\end{aligned}
$}
\label{eq:grpo}
\end{equation}
where $\epsilon$ and $\beta$ are hyperparameters. The KL term regularizes the policy toward a reference policy $\pi_{\mathrm{ref}}$. We apply the same retrieved-token masking and length normalization when computing $D_{\mathrm{KL}}$.

\section{Training Details}
\label{Training Details}
In this section, we provide the training details for reproducibility. We train Qwen3-8B using the verl framework on $8\times$ H200 GPUs. The training and validation batch sizes are 256 and 512, respectively, and we allow up to 8 interaction turns. We set the maximum sequence lengths to 16{,}384 tokens for the prompt and 8{,}192 tokens for the response. We use a learning rate of $1\times10^{-6}$ and set the entropy coefficient to 0. We use a PPO mini-batch size of 128 with a per-GPU micro-batch size of 32. For rollouts, we use vLLM with tensor parallelism set to 1, up to 128 concurrent sequences, and at most 131{,}072 batched tokens, sampling 8 responses per prompt. For both rollout and reference log-probability computation, we use a per-GPU micro-batch size of 32. We train for one epoch, run validation before training and every 20 steps thereafter, and save checkpoints every 20 steps. For all baselines, we follow the hyperparameters and training settings reported in the original papers. More details are provided in our released code.

\section{Process-level Reward Details}
\label{Process-level Reward Details}

In this section, we present the prompt used for computing the process-level reward in Figure~\ref{fig:prompt_scoring}.

\begin{figure*}[t]
\centering
\fcolorbox{black}{gray!10}{\parbox{0.99\textwidth}{
\textbf{Input}: \\
You are an experienced criminal judge. Given the case facts and a list of proposed sentencing factors, assess whether each factor can be established in this case based on judicial practice and common sense. Then provide an \emph{integer} total score from 1 to 10, where a higher score indicates that the list is overall more credible and that more factors are supported by the facts. \\
Factors may involve the determination of amounts/counts, modus operandi, and scenario (e.g., unlawful entry, pickpocketing, destructive means causing losses), identity and criminal history (e.g., recidivism), and post-offense or procedural factors (e.g., restitution, confession, guilty plea, plea acceptance). Reasonable inference is allowed, but you must not speculate beyond the facts. If the information is insufficient, assign a lower score. \\

\vspace{0.4em}
\textbf{Scoring rubric (1--10, integer)}: \\
9--10: Most factors clearly match the facts or can be supported with short-chain inference; key elements are covered with almost no conflicts. \\
7--8: Most factors are credible; a few details are missing or uncertain, but do not affect the overall assessment. \\
5--6: Only some factors are credible; multiple factors lack key elements and require additional information. \\
3--4: Few factors can be established; most are weakly inferred with insufficient factual support. \\
1--2: Key factors directly conflict with or are contradicted by the facts, or most factors are speculation without factual anchors. \\

\vspace{0.4em}
\textbf{Penalty triggers (decrease within the chosen band)}: \\
(1) Multiple factors are stated as generic labels, but the facts do not provide the required supporting elements: $-1$ to $-2$. \\
(2) A key factor directly conflicts with or is contradicted by the facts (e.g., unlawful entry, pickpocketing, recidivism not satisfied): $-3$ to $-5$. \\
(3) Factors are clearly speculative without factual anchors: $-1$ to $-3$. \\
(4) Factors are mutually inconsistent: $-1$ to $-3$. \\
The final score must remain within the range of 1--10. \\

\vspace{0.4em}
Output \textbf{only one integer} in the format \texttt{<answer>7</answer>}.\\
Case Facts: \{\ fact\ \} \\
Sentencing Factors: \{\ factors\ \} \\

\textbf{Output}: \\
\texttt{<answer>}\{\ score\ \}\texttt{</answer>}
}}
\caption{Prompt used for scoring whether the proposed sentencing factors are supported by the case facts.}
\label{fig:prompt_scoring}
\end{figure*}

\section{Case Study Details}
\label{Case Study Details}
\begin{figure*}[t]
    \centering
    \includegraphics[width=\textwidth]{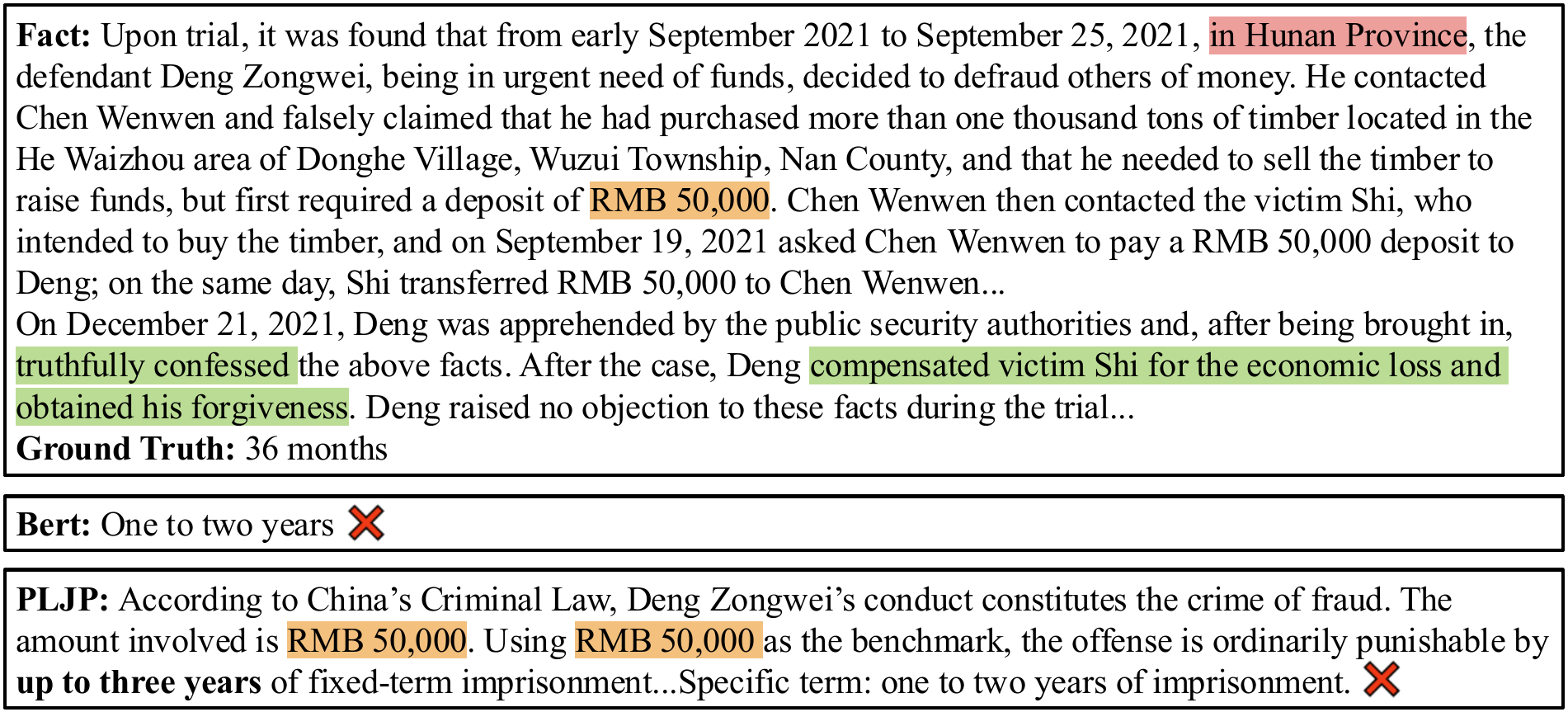}\par\medskip
    \includegraphics[width=\textwidth]{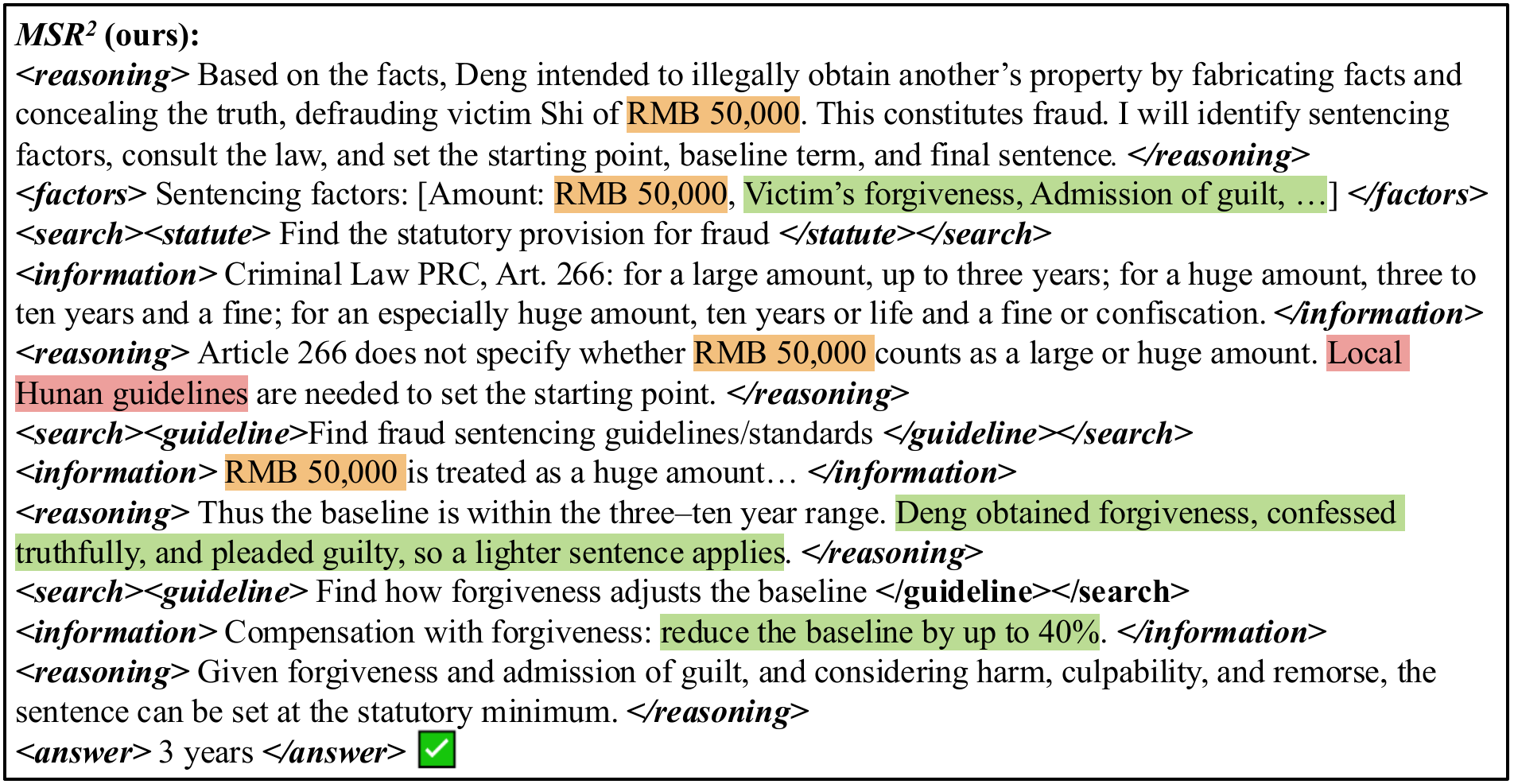}
    \caption{Case study of a fraud case. $MSR^2$ demonstrates a transparent decision process by selectively routing queries to appropriate sources and performing sound subjective reasoning by integrating retrieved information with case facts. The model correctly anchors the baseline term by verifying the specific amount threshold and subsequently adjusts the sentence based on mitigation factors, forming a coherent chain from facts to judgment.}
    \label{fig:3}
\end{figure*}

In this section, we present a qualitative case study of a fraud case in Figure~\ref{fig:3}.

\section{Expert Annotation Guidelines}
\label{app:expert_annotation}

To assess the reliability of the process-level rewards derived from LLM-as-a-judge, we conduct a human--model agreement study.

\noindent\textbf{Annotation Scope:}
We randomly sample 200 cases from the evaluation set for human annotation. For each case, the sentencing factors are first extracted by the policy model. The annotation task is not to re-predict the final sentence, but to assess whether the extracted factors are legally and factually valid. Each instance contains the case fact description and a list of proposed sentencing factors. These
factors may include objective elements, such as the amount involved, number of
offenses, offense type, and statutory circumstances, as well as subjective or
discretionary elements, such as confession, remorse, victim forgiveness,
self-surrender, intent, public-place determination, or other mitigating and
aggravating circumstances.

\noindent\textbf{Expert Annotation Procedure:}
Each sampled case is independently annotated by three legal experts from leading
law schools. Annotators are asked to evaluate the proposed sentencing-factor list on a 1--10 scale. A higher score indicates that the factor list is overall
more credible, more complete, and better supported by the case facts and common
judicial practice. A lower score indicates that the list contains unsupported,
irrelevant, or misleading factors, or omits important factors.

\noindent\textbf{Annotation Guidelines:}
Annotators are provided with a rubric that specifies the following criteria.
First, \textit{factual support} measures whether each factor is explicitly
stated in, or can be reasonably inferred from, the case facts. Factors that
contradict the facts or introduce unsupported assumptions should be penalized.
Second, \textit{legal relevance} measures whether the factor is relevant to
criminal sentencing under judicial practice, such as statutory sentencing
ranges, amount thresholds, mitigating circumstances, and aggravating
circumstances. Third, \textit{factor coverage} measures whether the factor list
captures the major sentencing circumstances necessary for determining the
sentence. Missing important factors should reduce the score even if the listed
factors are correct. Fourth, \textit{reasoning credibility} measures whether the
factors are stated clearly and consistently, without conflating legal concepts
or over-interpreting ambiguous facts. The final annotation score reflects the
overall quality of the factor list under these criteria.

\noindent\textbf{Score Aggregation:}
After independent annotation, we aggregate the three expert ratings for each
case. When a majority score exists, we use majority voting to determine the
final human annotation. When no majority exists, we use the median score as the
aggregated expert judgment.

\noindent\textbf{Review and Quality Control:}
Before formal annotation, annotators are provided with the task definition,
rubric, and representative examples to ensure a shared understanding of the
evaluation criteria. During annotation, experts evaluate each case
independently and are asked not to discuss their scores with other annotators.
After annotation, we conduct consistency checks to identify invalid entries,
missing scores, or cases where the score is inconsistent with the written
rationale. Ambiguous cases are reviewed by the authors, and annotation
instructions are clarified when necessary. These procedures help ensure that
expert judgments are reliable and comparable across cases.

\noindent\textbf{Feedback Mechanism:}
We maintain an iterative feedback mechanism during the annotation process.
Annotators are encouraged to report unclear criteria, ambiguous legal concepts,
or cases where the provided facts are insufficient to determine a factor. These
reports are reviewed and used to refine the annotation rubric, clarify decision
rules, and improve the consistency of subsequent annotations. This mechanism is
particularly important for legal sentencing prediction, where many factors
require both legal knowledge and case-specific reasoning.

\noindent\textbf{Annotator Compensation:}
All annotators are fairly compensated for their work according to the expected
annotation time and the expertise required for legal judgment. The compensation
covers case reading, factor assessment, scoring, and feedback on ambiguous
instances.

\noindent\textbf{Ethical Considerations:}
All annotators are informed that their annotations will be used for academic
research, and they provide consent before participating.
Annotators are asked to make judgments based only on the provided case facts,
the proposed sentencing factors, and the annotation rubric, rather than on any irrelevant information.

%% file: relate_work.tex
\label{app:related_work_details}
\subsection{Legal Judgment Prediction}

Legal Judgment Prediction (LJP) is a core task in legal AI that aims to predict judicial outcomes, such as applicable law articles, charges, and sentencing terms, by analyzing case facts with traditional models or large language models (LLMs). In this work, we broadly categorize existing LJP approaches into two paradigms: the semantic-focused paradigm and the logic–focused paradigm.

\subsubsection{Semantic-focused paradigm.}
This line of work typically advances along two directions. (1) Representation learning, which leverages traditional machine learning or neural networks to learn informative representations of case facts for classification. For example, LADAN~\cite{xu2020distinguish} addresses confusable law articles by explicitly modeling inter-article relations and learning discriminative features that capture subtle distinctions in case facts. NeurJudge~\cite{yue2021neurjudge} further emphasizes circumstance-aware modeling by using intermediate subtask predictions to partition fact descriptions into different crime circumstances and then deriving circumstance-specific representations to improve downstream predictions. To enhance class separability under fine-grained inter-class similarity, contrastive learning has also been adopted to explicitly separate hard positive and hard negative pairs, thereby sharpening decision boundaries for confusing outcomes~\cite{zhang2023contrastive,gan2023exploiting}. 
(2) Retrieval augmentation, which incorporates external information to enrich the factual context and improve decision accuracy. For instance, PLJP~\cite{wu2023precedent} follows this direction by combining LLMs with domain-specific models: the domain models efficiently retrieve relevant candidate precedents, and the LLM then makes context-aware decisions conditioned on the retrieved cases. Despite these advances, approaches that rely primarily on textual semantic similarity often fall short of the logical rigor required for sentencing. Moreover, they may overlook fine-grained legal sources, such as sentencing guidelines and judicial interpretations, which are essential for handling nuanced legal issues and producing reliable sentencing predictions.

\subsubsection{Logic-focused paradigm.}
This paradigm emphasizes that accurate legal judgments require logical reasoning that goes beyond simple textual semantic similarity. Early work like TopJudge~\cite{zhong2018legal} explicitly models the dependencies among subtasks, including law article prediction, charge prediction, and sentencing prediction, to better capture inter-task interactions. Other approaches formalize legal knowledge as logical constraints to guide learning. For example, \citet{gan2021judgment} encode declarative legal knowledge as first-order logic rules and integrate them into a neural architecture, enabling more explicit reasoning with identifiable legal bases. With the rise of LLMs, prompting-based and workflow-based reasoning has become increasingly popular. Legal syllogism prompting (LoT)~\cite{jiang2023legal} teaches LLMs to follow the classical legal syllogism, where the major premise is the law, the minor premise is the fact, and the conclusion is the judgment, to elicit deductive reasoning that aligns with legal analysis in case judgment prediction. Similarly, Ask-Discriminate-Predict (ADAPT)~\cite{deng2024enabling} introduces a reasoning framework inspired by human judicial thinking, where case facts are first decomposed, then potential charges are discriminated, and finally the judgment is predicted. Despite their improved logical grounding and interpretability, these logic-focused methods remain limited: they often capture only superficial topological structures, rely on rigid formalisms such as first-order predicate logic with restricted expressiveness, or depend solely on the intrinsic capabilities of LLMs, which constrain both the breadth and depth of inference. As a result, they may struggle to adapt to case-specific contradictions and to integrate external legal knowledge, such as sentencing guidelines and judicial interpretations, into reasoning, which is essential for nuanced and reliable sentencing predictions.

\subsection{RL-Driven Reasoning and Retrieval}
Recent advances in LLM reasoning have been driven by both inference-time techniques and training-time optimization~\cite{jaech2024openai}. Chain-of-thought prompting shows that explicitly eliciting intermediate reasoning steps can substantially improve performance on complex reasoning tasks~\cite{wei2022chain}. Moreover, DeepSeek-R1~\cite{guo2025deepseek} demonstrates that reinforcement learning with outcome-driven rewards can significantly encourage effective reasoning behaviors. Search-R1~\cite{jin2025search} extends this idea by training models to interleave step-by-step reasoning with search, enabling multi-turn query generation and retrieval-conditioned reasoning. More recently, DeepResearcher~\cite{zheng2025deepresearcher} explores end-to-end reinforcement learning for research agents in real-world environments.

Building on these insights, we propose $MSR^2$, a reinforcement learning–driven framework for legal sentencing prediction that introduces two key advances: a multi-source retrieval module and a process-level reward. By tightly integrating multi-source retrieval with reasoning, $MSR^2$ enables broader and deeper inference than prior approaches. We believe that our framework provides useful insights for general reasoning and retrieval tasks beyond the legal domain.